\documentclass[12pt]{article}

\usepackage{xspace}
\usepackage{amstext}
\usepackage{enumerate}
\usepackage{citesort}
\usepackage[mathscr]{eucal}
\usepackage{epsfig}
\usepackage{amsmath}
\usepackage{theorem}
\usepackage{psfig}
\usepackage{graphics}
\usepackage{psfrag}

\def\myendproof{{\ \vbox{\hrule\hbox{%
   \vrule height1.3ex\hskip0.8ex\vrule}\hrule }}\par}
 \newtheorem{theorem}{Theorem}[section]
\newtheorem{lemma}[theorem]{Lemma}
\newtheorem{corollary}[theorem]{Corollary}
\newtheorem{fact}[theorem]{Fact}

\newtheorem{observation}[theorem]{Observation}
\newenvironment{proof}{{\it Proof. }}{\myendproof}

\newcommand{\comment}[1]{}

\newcommand{\nni}{{NNI}\xspace}
\newcommand{\stt}{{STT}\xspace}
\newcommand{\sttdist}{{\tt STTdist}}
\newcommand{\rsttdist}{{\tt rSTTdist}}

\title{\vspace*{-.3in}
Improved Phylogeny Comparisons: Non-Shared Edges,
Nearest Neighbor Interchanges, and Subtree Transfers
\footnote{A preliminary version appeared in
the Proceedings of the 11th International Conference on
Algorithms and Computation (ISAAC), Lecture Notes in Computer Science 1969,
Springer, Taipei, Taiwan, 527--538, 2000.}
}

\author{%
        Wing-Kai Hon\thanks{Department of Computer Science, University
of Hong Kong, Hong Kong. Email: \{wkhon,twlam,smyiu\}@csis.hku.hk. }
	\and
	Ming-Yang Kao\thanks{Department of Computer Science,
Northwestern University, Evanston, IL 60201, U.S.A. Email: kao@cs.northwestern.edu.
Supported in part by NSF grant EIA-0112934.}
        \and
        Tak-Wah Lam\footnotemark[1]
	\and
	Wing-Kin Sung\thanks{Department of Computer Science,
National University of Singapore, Singapore. Email: ksung@comp.nus.edu.sg. }
	\and
	Siu-Ming Yiu\footnotemark[1]
}

\begin{document}
\maketitle
\thispagestyle{empty}

\begin{abstract}
The number of the non-shared edges of two phylogenies is a basic measure of the dissimilarity
between the phylogenies. The non-shared edges are also the building block
for approximating a more sophisticated metric called 
the nearest neighbor interchange (NNI) distance.
In this paper, we give the first subquadratic-time algorithm for finding the non-shared
edges, which are then used to speed up the existing approximating algorithm for
the NNI distance from $O(n^2)$ time to $O(n \log n)$ time.
Another popular distance metric for phylogenies is 
the subtree transfer (STT) distance.
Previous work on computing the STT distance considered degree-$3$ trees only.
We give an approximation algorithm for the STT distance for degree-$d$
trees with arbitrary $d$ and with generalized STT operations.
\end{abstract}

\section{Introduction}
{\it Phylogenies} are trees whose leaves are labeled with distinct 
species.  
Different theories about the evolutionary relationship 
of the same species often result in 
different phylogenies.
This paper is concerned with three well-known metrics for
measuring the dissimilarity between two phylogenies,
namely, the {\it non-shared edge} distance
\cite{Bourque:1978:ADS,Robinson:1981:CPT,Li:1996:SNN},
the {\it nearest neighbor interchange}(NNI) distance
\cite{Robinson:1971:CLT,Moore:1973:AIA}
and the {\it subtree transfer}(STT) distance \cite{Hein:1990:RES,Hein:1993:HMR}.
The first metric counts the number of edges that differentiate
the phylogenies;
the other two metrics measure the minimum total cost of 
some kind of tree operations
required to transform one phylogeny to the other.
For the \nni distance, an operation swaps
two subtrees over an internal edge; for the \stt distance,
an operation detaches a subtree from a node
and re-attaches it to another part of the tree.

In this paper we consider phylogenies of degrees $d$
whose edges may carry weights.\footnote{%
	The magnitude of an edge weight, also known as
	branch length in genetics,
	may represent the number of mutations or the
	time required by the evolution along the edge.}
Given two weighted degree-$d$ phylogenies $T$ and $T'$, 
an edge $e$ in $T$ is {\em shared}\/ if for
some edge $e'$ in $T'$, the removals of $e$ and $e'$ from $T$ and $T'$, respectively
induce the same
partition of leaf labels, internal node degrees, and edge weights;
otherwise, $e$ is {\em non-shared}.
Previously, non-shared edges could
be found using a brute-force approach in $O(n^2)$ time,
where $n$ is the number of leaves.
If we restrict our attention to the partition of leaf labels only,
Day  \cite{Day:1985:OAC}
reduced the time to $O(n)$.
We give an $O(n \log n)$-time algorithm
for finding the general non-shared edges.

Finding non-shared edges is a key step,
as well as the most time consuming step,
 for approximating the \nni distance.
In particular, for degree-3 phylogenies with weights or
degree-$d$ phylogenies with or without weights, 
existing approximation algorithms take $O(n^2)$ time 
\cite{DasGupta:1997:ODB,Hon:1999:ANN}.
With our new non-shared edge algorithm, the time complexity of
these approximation algorithms can all be  improved to $O(n\log n)$.
Note that for unweighted degree-3 trees, 
an $O(n \log n)$-time algorithm has already been obtained
\cite{Li:1996:SNN}, which uses
Day's linear-time algorithm \cite{Day:1985:OAC} to identify the non-shared 
edges.

\begin{figure}[t]
\centerline{\psfig{figure=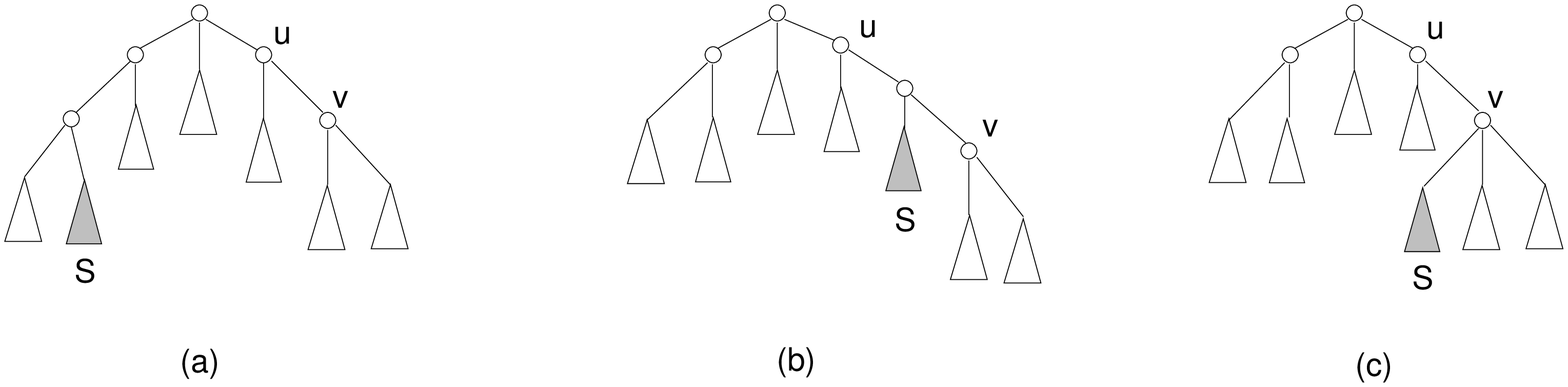,height=1.25in,width=0.98\textwidth}}
\caption{Examples of restricted-\stt and \stt operations. Tree (a) is 
transformed to tree (b) by a restricted-\stt operation, while tree(a) is
transformed to tree (c) by an \stt operation.} \label{stt_example1}
\end{figure}

Previous work on \stt distance focuses
on degree-3 trees only \cite{Hein:1996:OCC,DasGupta:1999:OLC}.  
In particular, in the course of transforming a degree-3
tree to another degree-3 tree,
all intermediate trees are required to be of degree 3.  In other words,
the \stt operation is {\em restricted} in the sense that
the subtree detached can only be re-attached to the middle of an edge, 
producing a new internal node with degree 3.  
See Figure~\ref{stt_example1} for an example.  In this
paper we study the \stt distance for
degree-$d$ phylogenies for any $d \geq 3$ while also
allowing an \stt operation to re-attach
the subtree to either an internal node or the middle of an edge.

\begin{figure}
\centerline{\psfig{figure=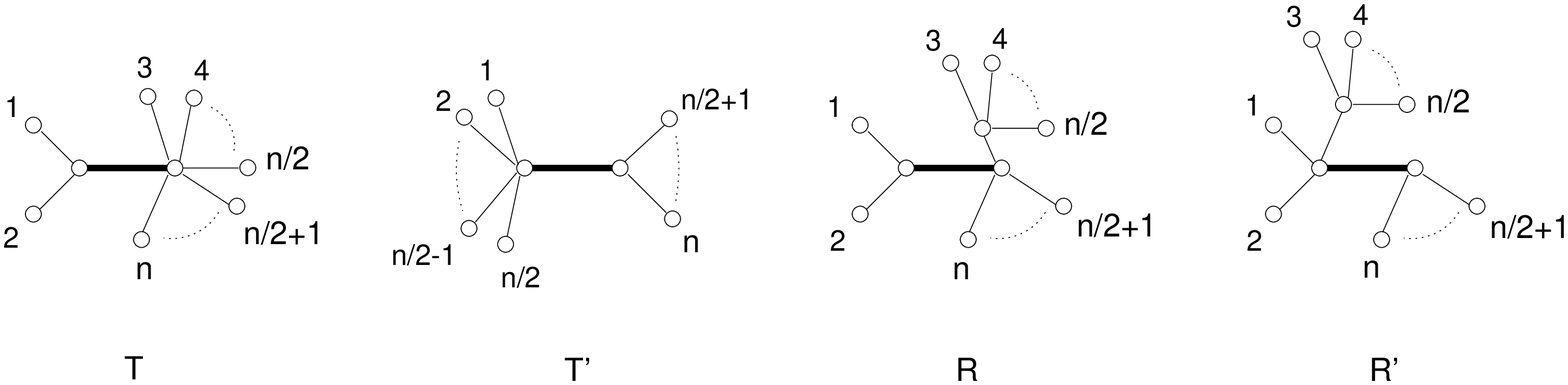,height=1.25in,width=0.98\textwidth}}
\caption{The unweighted STT distance between $T$ and $T'$ is
$\Omega(n)$.  Consider $T$ and $T'$ as weighted
trees such that every internal edge has a unit weight
(i.e., the highlighted edges in the figure).
The weighted STT distance between $T$ and $T'$ is 1.  In particular,
the cost of transforming $T$ to $R$, then to $R'$,
and finally to $T'$ is $0 + 1 + 0 = 1$.}
\label{stt_example2}
\end{figure}

\begin{figure}[ht]
\centerline{\psfig{figure=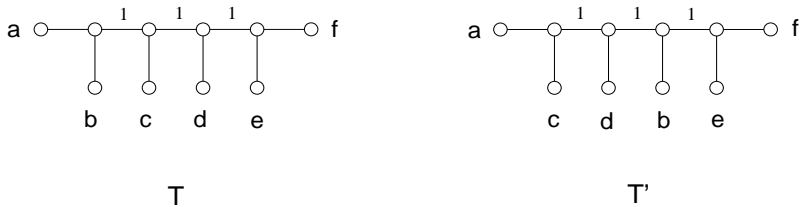,height=1.05in,width=0.65\textwidth}}
\caption{$T$ and $T'$ are degree-3 phylogenies. The unweighted STT
distance between $T$ and $T'$ (which is 3) is smaller than the weighted
STT distance (which is 4).}
\label{stt_example3}
\end{figure}

An \stt operation is charged by how far
a subtree is transferred.  More specifically,
depending on whether the trees are unweighted or weighted,
we count respectively the number of the edges or 
the total weight of the edges\footnote{%
This cost model is referred to as the linear cost model
in the literature \cite{DasGupta:1997:ODB,DasGupta:1999:OLC}.  
It is preferable to the unit cost model
as the latter does not reflect the evolutionary distance.}
between the nodes where detachment and re-attachment take place.
We formally define the \stt (respectively, {\em restricted-\stt}) distance
between two phylogenies
as the minimum cost of transforming one to the other
using \stt (respectively, restricted-\stt) operations.
Unlike many other graph or tree problems,
the unweighted version of the \stt distance problem
is not a special case of the weighted version.
In particular, Figure~\ref{stt_example2} shows two phylogenies 
whose unweighted \stt distance is $\Omega(n)$,
yet if we assign a unit weight to every edge of these phylogenies,
their weighted \stt distance is only $O(1)$.
On the other hand, the unweighted \stt distance is not necessarily
bigger than the weighted one; Figure~\ref{stt_example3} shows two
phylogenies whose unweighted \stt distance is indeed smaller than the
weighted one.

Consider degree-3 phylogenies. In the weighted case,
we can prove that
the \stt distance is the same as the restricted-\stt distance,
and DasGupta et al.\ have shown that
the latter can be approximated within a factor of 2 \cite{DasGupta:1999:OLC}.
In the unweighted case, deriving a tight approximation algorithm
is more difficult; the restricted-\stt distance can be approximated within
only a factor of $O(\log n)$ \cite{DasGupta:1999:OLC}.
This result implies an approximation algorithm for the \stt distance
 with the same performance. 
However, there are examples in which
the \stt distance is much smaller than the restricted-\stt distance.
It is natural to ask whether the \stt distance can be approximated
within a better factor.

Consider degree-$d$ phylogenies.
First of all, it is worth mentioning that the restricted-\stt
distance is $\infty$ as a restricted-\stt operation can
only produce an internal node of degree 3.
In the weighted case, the \stt distance can be approximated
by adapting the algorithm by DasGupta et al for degree-3 trees
\cite{DasGupta:1999:OLC}, achieving the approximation factor of 2.
In the unweighted case,
we give an algorithm to approximate the \stt distance
within a factor of $2d- 4$.  
This result implies that for unweighted degree-3 trees, the
approximation factor can be improved from $O(\log n)$ to $2$
if the intermediate trees may not be necessarily degree-$3$ trees.
Table~\ref{table-1} summarizes the approximation factors
for the variants of the \stt distance.

\begin{table}
\begin{center}
\caption{The approximation factors
for the different variants of the \stt distance. \vspace{1ex}}
\begin{tabular}{||l|c|c||}
	\hline
 & restricted-\stt  & \stt \\ 
 & (degree-3 trees) & (degree-$d$ trees) \\ \hline
 weighted	& 2 \cite{DasGupta:1999:OLC} & 2 \\ \hline
 unweighted	& $O(\log n)$ \cite{DasGupta:1999:OLC} & $2d -4$ \\ \hline
\end{tabular}
\label{table-1}
\end{center}
\end{table}

The following summarizes the contributions of the paper.
\begin{enumerate}
\item We give the first subquadratic (i.e., $O(n \log n)$) time
      algorithm for finding the non-shared edges between two
      weighted phylogenies, thus improving
      the time complexity of the algorithms in
      \cite{DasGupta:1997:ODB,Hon:1999:ANN}
      for approximating the NNI distance
      from $O(n^2)$ to $O(n \log n)$.
\item We show that the problem of finding the STT distance between
      two weighted degree-$d$ phylogenies is equivalent to
      the problem of finding the restricted-STT distance between
      two weighted degree-3 phylogenies. This result implies the
      following.
      \begin{enumerate}
      \item The STT distance between two weighted degree-$d$
            phylogenies can be approximated within a factor of 2.
      \item If the leaf labels of the trees are not distinct, it is
            NP-hard to compute the STT distance between two trees.
      \end{enumerate}
\item We give an approximation algorithm with approximation ratio
      of $2d-4$ for finding the STT distance between two unweighted
      degree-$d$ phylogenies.
\item We prove that it is NP-hard to compute the STT distance between
      two unweighted trees with leaves labeled by possibly non-distinct
      labels.
\end{enumerate}

The rest of this paper is organized as follows.
Section~\ref{sec-non-shared-edges} gives the $O(n \log n)$-time algorithm for finding
the non-shared edges between two weighted trees.
Section~\ref{sec-stt} presents the results on 
computing subtree transfer distance for both weighted and unweighted cases.
Finally, Section~\ref{sec-STT-nphard} shows that computing STT distance
between two unweighted phylogenies with possibly non-distinct labels is NP-hard.

\section{Finding non-shared edges between weighted trees}
\label{sec-non-shared-edges}
In this section, we show how to find all non-shared
edges between two weighted phylogenies in $O(n \log n)$ time.
Basically, we tranform the problem to a partition labeling
problem which can be solved in $O(n \log n)$ time.
In Section \ref{sec-partition-labeling}, we
define the partition labeling problem on two rooted trees and 
solve the problem in $O(n \log n)$ time. In Section
\ref{sec-alg-nonsharededges},
we define the non-shared edge problem and give an $O(n \log n)$-time
reduction from the non-shared edge problem to the partition
labeling problem.

The following multi-set notations are used in this section.
Let $M = \{a,a,\dots, b,b,\dots\}$ be a multi-set of symbols.
Let $\delta(M)$ be the set of distinct symbols in $M$.
For each $a \in \delta(M)$, let $\#a$ be the number of
occurrences of $a$ in $M$. 
Let $|M| = \sum_{a \in \delta(M)}{\#a}$.
Furthermore, the set operations for any two multi-sets $M$ and
$N$ are defined as follows: 
(i) $M \subseteq N$ if
for each $a \in \delta(M)$, $\#a$ in $M \leq \#a$ in $N$. 
(ii) $M = N$ if $M \subseteq N$ and $N \subseteq M$. 
(iii) $M \cup N$ is defined to be a multi-set containing
$\max\{\#a \mbox{~in~} M, \#a \mbox{~in~} N\}$ copies of $a$
for each $a$ in $\delta(M) \cup \delta(N)$.

\subsection{The partition labeling problem for rooted trees}
\label{sec-partition-labeling}
In this section we define the partition labeling problem
for two rooted trees with leaves labeled by
the same multi-set of labels and solve the problem
in $O(n \log n)$ time, where $n$ is the number
of leaves in either tree. In the following,
let $R$ and $R'$ be two rooted trees
with leaves labeled by the same multi-set $S$ of labels, and let
$A$ be any subset of $\delta(S)$.

For each internal node $u$ in $R$, we define the following.
\begin{itemize}
\item Let $L_R(u)$ be the multi-set of leaf labels in the subtree of
$R$ rooted at $u$.
\item Let $L_R(u)|A$ be
the multi-set of leaf labels constructed from
$L_R(u)$ by deleting all labels which are not in $A$.
\end{itemize}

Given $R$ and $R'$, let $V$ be the union of the sets of
internal nodes in $R$ and $R'$. A mapping
$\rho: V \rightarrow [1..t]$ (where $t = |V|$) is called
a partition labeling for $R$ and $R'$ if for any
$u$ in $R$ and $v$ in $R'$, $\rho(u) = \rho(v)$
if and only if $L_R(u) = L_{R'}(v)$.

The {\it partition labeling} problem is to find a partition labeling
for $R$ and $R'$.
Note that this partition labeling always exists.
A straightforward approach is to compute all multi-sets of $L_R(u)$
(and $L_{R'}(v)$) first and then assign a unique integer to each distinct
multi-set.  However, each multi-set can be as large as $O(n)$, so this
straightforward approach takes $O(n^2)$ time.

To reduce the time complexity,  we compute the multi-sets in an
incremental manner, and start comparing them earlier based on the
partial result.  In particular, we assign a temporary label to represent
each multi-set, such that two multi-sets are assigned the same label if
they are equal.  This helps in saving not only the space for storing
the multi-sets, but also the time for comparing two multi-sets
afterwards.  The labels will further be updated by a relabeling process, 
as long as more information about the multi-sets are computed.  In
the end, each distinct multi-set of $L_R(u)$ (and $L_{R'}(u)$) obtains a
distinct label.

The algorithm is presented in Figure~\ref{fig-partition-labeling}, where
$R_A$ is defined as the subtree of $R$ induced by $A$.  Precisely, $R_A$
is a tree constructed by contracting $R$ to retain only those leaves with
labels in $A$ and their least common ancestors.

The algorithm is analysed below, and we begin with two supporting lemmas.

\begin{figure}[htbp]
\fbox{
\begin{minipage}{.95\textwidth}
\noindent
\begin{description}
  \item[Phase 1.]  For each label $i$ in $\delta(S)$,
             find a
             parition labeling for $R_{\{i\}}$ and $R'_{\{i\}}$ according
             to Lemma \ref{basis}.
  \item[Phase 2.]  Repeat the following procedure for $\log |\delta(S)|$
             rounds: \\
             Let $A_1, A_2, \ldots$ be the sets of labels considered
             in the last step.
             \begin{enumerate}
             \item Pair up $A_i$'s such that $A_{2j-1} = A_{2j-1} \cup A_{2j}$. 
             \item Delete all $A_{2j}$'s and rename $A_{2j-1}$ as $A_j$. 
             \item For each $A_j$, compute a partition labeling for $R_{A_j}$ and 
             $R'_{A_j}$ based on the result of the last step
             and Lemma \ref{LabelsofRAUB}.
             \end{enumerate}
\end{description}
\end{minipage}}
\caption{Partition labeling for $R$ and $R'$.}
\label{fig-partition-labeling}
\end{figure}

\begin{lemma}
The induced subtree $R_A$ can be constructed
in $O(t)$ time where $t$ is the number of leaves in $R$ with
labels in $A$.
\end{lemma}
\begin{proof}
Using the algorithm in \cite{Harel:1984:FAF}, with linear time preprocessing,
we can answer a least common ancestor query in constant time.
To construct $R_A$, we only need to answer $O(t)$ least common
ancestor queries where $t$ is the number of leaves in $R$ with
labels in $A$, so the lemma follows.
\end{proof}

\begin{lemma}
\label{findLRAUB}
Let $A$ and $B$ be two disjoint subsets of $\delta(S)$.
Let $u$ be an internal node in $R_{A \cup B}$. Then
$L_{R_{A \cup B}}(u)|A = \emptyset$ or
$L_{R_A}(v)$ for some $v$ in $R_A$.
(Similarily, $L_{R_{A \cup B}}(u)|B = \emptyset$ or
$L_{R_B}(v)$ for some $v$ in $R_B$.)
\end{lemma}
\begin{proof}
If $u$ is in $R_A$, $L_{R_{A \cup B}}(u)|A = L_{R_A}(u)$.
It remains to consider $u$ not in $R_A$. Suppose on the
contrary that
$L_{R_{A \cup B}}(u)|A \neq \emptyset$
and $L_{R_{A \cup B}}(u)|A \neq L_{R_A}(v)$
for any internal node $v$ in $R_A$. Let this $u$ be the first
one of this kind visited by a postorder
traversal of $R_{A\cup B}$.

By the construction of $R_A$, since $u$ is not in $R_A$,
$u$ has at most one child $s$ whose subtree contains
leaf labels in $A$.
If such an $s$ exists, then $L_{R_{A \cup B}}(u)|A = L_{R_{A \cup B}}(s)|A$.
This contradicts the choice of $u$.  If $u$ has no such a child, 
$L_{R_{A \cup B}}(u)|A = \emptyset$.
Thus such a $u$ does not exist. A similar argument can be applied to
the case of $L_{R_{A \cup B}}|B$.
\end{proof}

The next lemma implies that
Phase 1 can be computed in $O(n)$ time.
\begin{lemma}
\label{basis}
Let $a \in \delta(S)$. A partition labeling for
$R_{\{a\}}$ and $R'_{\{a\}}$
can be found in $O(t)$ time where $t$ is the number of
leaves in $R$ with label $a$.
\end{lemma}
\begin{proof}
Perform a postorder traversal in $R_{\{a\}}$.
Since $L_{R_{\{a\}}}(u)$ only contains multiple copies of $a$, we
only need to keep track $|L_{R_{\{a\}}}(u)|$ during the
traversal and assign this number to the internal node $u$.
Apply the same procedure to $R'_{\{a\}}$. 
\end{proof}

After Phase 1, we get the partition labeling for $R_{\{i\}}$ for every
label $i \in \delta(S)$.
Phase 2 tries to merge the $R_{\{i\}}$'s incrementally until we
get the partition labeling for $R_{\delta(S)}$.
Below we describe the merging process.
Let $A$ and $B$ be two disjoint subsets of $\delta(S)$. Let $\rho_1$
and $\rho_2$ be partition labelings for $(R_A, R_A')$ and
$(R_B, R_B')$, respectively.  Now, we perform the following
relabeling process on the internal nodes of 
$R_{A \cup B}$ and $R'_{A \cup B}$.

\vspace{5pt}
\noindent{\bf Relabeling Process:}
Consider $R_{A \cup B}$ (similar for $R'_{A \cup B}$).
\begin{description}
\item[Step 1.] {Perform a postorder traversal.
          For each internal node $u$ in $R_{A \cup B}$ visited,
          assign a 2-tuple of integers $(a, b)$ to $u$ in the
          following manner: \\
          According to Lemma \ref{findLRAUB},
          if $L_{R_{A \cup B}}|A = \emptyset$, set $a$ to 0;
          Otherwise, there exists a $v$ such that 
          $L_{R_{A \cup B}}|A = L_{R_A}(v)$, then set $a$ to $\rho_1(v)$.
          Set $b$ similarily by considering $L_{R_{A \cup B}}|B$.}
\item[Step 2.] Sort all 2-tuples of internal nodes by radix sort. Traverse
          the sorted list of these 2-tuples, assign a new integer
          (starting from 1) to every distinct 2-tuple encountered.
          Assign this integer as a label to the corresponding internal
          node.
\end{description}

In fact, the labels assigned to the nodes in the relabeling
process form a valid partition labeling. We have the following
lemma.

\begin{lemma}
\label{LabelsofRAUB}
Given the partition labelings for $(R_A, R_A')$ and
$(R_B, R_B')$, we can compute a partition labeling for
$R_{A \cup B}$ and $R'_{A \cup B}$ in $O(t)$ time
where $t$ is the number of leaves in $R_{A \cup B}$.
\end{lemma}
\begin{proof}
Perform the relabeling process on $R_{A \cup B}$.
Since $L_{R_{A \cup B}}(u) = L_{R_{A \cup B}}(u)|A \cup
L_{R_{A \cup B}}(u)|B$, $L_{R_{A \cup B}}(p)
= L_{R'_{A \cup B}}(q)$ if and only if the corresponding
2-tuples assigned to $p$ and $q$ are identical. So, the
labels assigned to the nodes in Step 2 form a valid
parition labeling.

Regarding the time complexity, in Step 1, during the
postorder traversal,
for each internal node $u$, if $u$ is in $R_A$,
then $v = u$. Otherwise, if there exists a child $s$ of $u$
in $R_{A \cup B}$ with $L_{R_{A \cup B}}(s)|A = L_{R_A}(t)$,
then $v = t$. If no such $s$ exists,
$L_{R_{A \cup B}}(u)|A = \emptyset$. So, this Step can be
completed in linear time.
Obviously, Step 2 can also be completed in linear time, so the
lemma follows.
\end{proof}

Lemma~\ref{LabelsofRAUB} implies that each round of Phase 2
takes $O(n)$ time.
This gives the following lemma.

\begin{lemma}
Given $R$ and $R'$, a partition labeling for $R$ and $R'$
can be computed in $O(n \log n)$ time where $n$ is the number
of leaves in $R$.
\end{lemma}
\begin{proof}
By Lemma \ref{basis}, Phase 1 takes $O(n)$ time.
By Lemma \ref{LabelsofRAUB}, each round in Phase 2 takes
$O(n)$ time, so the overall complexity is
$O(n \log |\delta(S)|) = O(n \log n)$ time. 
Thus, the lemma follows.
\end{proof}

\subsection{An $O(n \log n)$-time algorithm for finding non-shared edges}
\label{sec-alg-nonsharededges}
In this section we show that the non-shared edge
problem for weighted phylogenies can be solved
by an $O(n \log n)$-time reduction to the partition labeling problem
on two rooted trees.

Let $T$ and $T'$ be two weighted phylogenies with
the same set of distinct leaf labels and the same multi-set of
edge weights and internal node degrees. Recall that a shared edge
is defined as follows.

An edge $e$ in $T$ is said to be
{\it shared} (with respect to $T'$) if there exists an edge $e'$ in $T'$ such
that $e$ and $e'$
induce the same partition of leaf labels, internal
node degrees, and edge weights in $T$ and $T'$, respectively;
otherwise, $e$ is {\it non-shared}.

The {\it non-shared edge}
problem is to find all non-shared edges in $T$ (with respect to $T'$)
and all non-shared edges in $T'$ (with respect to $T$).
Figure \ref{Transformation} gives the details of the reduction from
the non-shared edge problem to the partition labeling problem.
Basically, edge weights and node degrees in $T$ and $T'$ will
be represented by new labeled leaves in the
two constructed trees $R$ and $R'$, respectively.

\begin{figure}[htbp]
\fbox{
\begin{minipage}{.95\textwidth}
\noindent
\begin{itemize}
  \item[1.] Set $R = T$ and $R' = T'$.
  \item[2.] Fix an arbitrary leaf with label $a$. Root $R$ and $R'$
            at the internal nodes which attach to leaves with label $a$.
  \item[3.] Attach a new leaf to every internal node of $R$ and $R'$
            such that the labels of such new leaves are the same if the
            corresponding internal nodes have the same degree.
  \item[4.] Attach a new leaf in the middle of every internal edge
            in both $R$ and $R'$ such that the labels of
            such new leaves are the same if the original edges
            have the same weight.
\end{itemize}
\end{minipage}
} 
\caption{Construction of $R$ and $R'$ from $T$ and $T'$.}
\label{Transformation}
\end{figure}

Note that $R$ and $R'$ have the
same multi-set of leaf labels since $T$ and $T'$ have the same
multi-set of leaf labels, edge weights, and node degrees.
And the number of leaves in $R$ and $R'$ is of $O(n)$ where $n$
is the number of leaves in $T$.

\begin{lemma}
The construction of $R$ and $R'$ takes $O(n \log n)$ time.
\end{lemma}
\begin{proof}
The lemma follows since Step 4 takes at most $O(n \log n)$ time,
Step 3 takes $O(n)$ time, and Steps 1 and 2 take $O(1)$ time.
\end{proof}

The following lemma relates the non-shared edges problem and the
partition labeling problem.

\begin{lemma}
\label{reduction}
Given a partition
labeling $\rho$ for $R$ and $R'$, let $(u,v)$ be an edge in $T$
and $s$ be the unique internal
node between $u$ and $v$ in $R$. Then,
the edge $(u,v)$ is a non-shared edge in $T$ (w.r.t. $T'$)
if and only if the label $\rho(s)$ is unique in $\rho$.
\end{lemma}
\begin{proof}
Suppose that $(u, v)$ is a non-shared edge in $T$.
By the construction of $R$ and $R'$,
$L_R(s)$ must be unique. So, $\rho(s)$ is unique in $\rho$.

On the other hand, if $(u,v)$ is a shared-edge in $T$,
then there is another edge $(u',v')$ in $T'$ such that $(u, v)$
and $(u',v')$ induce the same partition of leaf labels, node degrees,
and edge weights in $T$ and $T'$, respectively. Without loss
of generality, let $u$ and $u'$ be
the portion containing the leaf with label $a$. 
Then, $u$ and $u'$ are the
ancestors of $v$ and $v'$ in $R$ and $R'$ respectively.
Let $s'$ be the unique internal node between $u'$ and $v'$.
By the construction of $R$ and $R'$, $L_R(s) = L_{R'}(s')$.
\end{proof}

In conclusion, we have the following theorem.

\begin{theorem}
The non-shared edges of $T$ and $T'$ can be identified 
in $O(n \log n)$ time.
\end{theorem}
\begin{proof}
By Lemma \ref{reduction}, if a partition labeling for $R$ and $R'$
is given, we can identify all non-shared edges in $T$ and $T'$ in
$O(n)$ time. Since the
parition labeling problem can be solved in $O(n \log n)$ time,
the theorem follows.
\end{proof}

\section{The \stt distance between degree-$d$ phylogenies}
\label{sec-stt}
This section studies the problem of computing the STT
distance between two degree-$d$ phylogenies in both weighted
and unweighted cases. 
For the weighted case, we show that the problem
of computing the STT distance between
two weighted degree-$d$ phylogenies (the weighted {\it STT-d} problem)
is equivalent
to the problem of computing the restricted-STT
distance between two weighted degree-$3$
phylogenies (the weighted {\it rSTT-3} problem). Since DasGupta et al
\cite{DasGupta:1999:OLC} have shown that the
weighted rSTT-3 problem
is NP-hard and
can be approximated within a factor of 2, the same
results apply to the weighted STT-$d$ problem.
For the unweighted case, we give a new approximation algorithm with
an approximation factor of $2d-4$ for finding the STT distance
between two degree-$d$ phylogenies (the unweighted STT-d problem). 
We also prove that the problem of
computing the STT distance between two unweighted phylogenies
with possibly non-distinct labels is NP-hard.

Section \ref{sec-preliminaries} gives notations and defintions used
in this section. 
Section \ref{sec-STT-weightedcase} gives
the result for the weighted case.
Section \ref{sec-approx-STT-unweightedcase} gives
an approximation algorithm 
for computing the STT distance between
two unweighted phylogenies.
Section \ref{sec-STT-nphard} shows that
it is NP-pard to compute STT distance between
two unweighted phylogenies with possibly non-distinct labels.

\subsection{Preliminaries} \label{sec-preliminaries}
Recall that \stt operation, restricted-STT operation,
STT distance and restricted-STT distance are defined as follows.
Given a tree $T$ (rooted or unrooted),
a subtree transfer (STT) operation is defined as
follows.
We select a subtree $S$ from $T$. Suppose $S$ is
attached to a node $u$
by an edge $e$. Pick another edge $e'=(v,w)$
(or an internal node $t$) not in $S$.
Detach $e$ and $S$ and re-attach them to a
newly created node $x$ in $e'$ (or $t$). If $u$ becomes degree 2
after removing $S$, merge the two edges connected to $u$ into one.
In the weighted version, let $w(e)$ denote the weight of an
edge $e$; we require that if a new node $x$ is created, then
$w((v,x))+w((x,w))=w(e')$; furthermore,
if $u$ is removed, the weight of the merged edge is
the total weight of the two merging edges.

An STT operation is called {\it restricted} if $S$ is always
re-attached to a new node inside an edge.
An STT operation is charged by how far the subtree is transferred.
Precisely, the cost of an STT operation
is defined as the number of edges or the total weight of edges, for
unweighted and weighted version, respectively,
on the shortest path from $u$ to $x$ (or $t$).

The STT distance between two trees $T_1$
and $T_2$, denoted by $\sttdist(T_1,T_2)$,
is defined as the minimum
cost of transforming $T_1$ to a tree which is leaf-label preserved
isomorphic to $T_2$ using STT
operations. The restricted-STT distance, denoted
by $\rsttdist(T_1,T_2)$, is defined
similarily by allowing only restricted-STT operations.

Note that $\sttdist(T_1,T_2)=\sttdist(T_2,T_1)$. However,
the corresponding equality may not hold for restricted-STT distance.
For example, consider the case that $T_1$ is a degree-4 tree 
while $T_2$ is a degree-3 tree.  
It is possible to transform $T_1$ to $T_2$ using restricted-STT
only operations, but not possible vice versa.

\subsection{Weighted degree-$d$ phylogenies}
\label{sec-STT-weightedcase}
This section shows that the problem of finding the STT
distance between two weighted degree-$d$ phylogenies
(the weighted STT-d problem) is
equivalent to the problem of finding the restricted-STT
distance between two weighted degree-$3$ phylogenies
(the weighted rSTT-3 problem).
We first show that the weighted STT-d problem can be reduced
to the weighted rSTT-3 problem. Given a weighted
degree-$d$ phylogeny $X$, we construct
a degree-$3$ phylogeny $T$ from $X$ as follows. 

\begin{figure}[htbp]
   \begin{center}   
   \psfrag{e0}{{$e_0$}} 
   \psfrag{e1}{{$e_1$}} \psfrag{y1}{{$y_1$}} 
   \psfrag{e2}{{$e_2$}} \psfrag{y2}{{$y_2$}} 
   \psfrag{ek-1}{{$e_{k-1}$}} 
   \psfrag{ek-2}{{$e_{k-2}$}} 
   \psfrag{ek-3}{{$e_{k-3}$}} 
   \psfrag{yk-3}{{$y_{k-3}$}} 
   \psfrag{u}{{$u$}} \psfrag{v}{{$v$}} 
   \psfrag{=>}{{\Large $\Rightarrow$}}      
   \resizebox{5in}{!}{\includegraphics{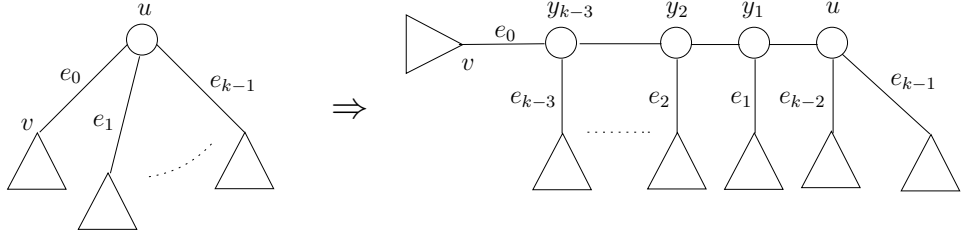}}
   \end{center}
\caption{Degree-$d$ node to Degree-$3$ node.}
\label{degd-deg3-trans}
\end{figure}

\vspace{5pt}
\noindent {\bf Transformation from a degree-$k$ phylogeny to a degree-$3$ phylogeny:}
For each node $u$ of $X$ with degree $k >3$, let
the edges that are attached to $u$ be
$e_{0}, e_{1}, \ldots, e_{k-1}$. Pick one of the
edges, say
$e_{0} = (u, v)$. Create
$k-3$ new nodes $y_1, y_2, \ldots, y_{k-3}$ on
$e_{0}$ such that $y_1$ is adjacent to $u$,
$y_i$ is adjacent to $y_{i-1}$ for $2 \leq i \leq k-3$
and $w((y_1, u)) = 0$, $w((y_i, y_{i-1}))= 0$ for
$2 \leq i \leq k-3$, $w((v, y_{k-3})) = w(e_{0})$.
Detach $e_i$ and the corresponding subtree, and reattach them
to node $y_i$ for $1 \leq i \leq k-3$. See figure \ref{degd-deg3-trans}
for an example.

Let $T$ be the resulting tree after the transformation.
Note that $T$ is a degree-$3$ phylogeny and 
the transformation only uses restricted-STT
operations of zero cost. 
We call $T$ or any tree which can be generated by the transformation
a degree-3 representation of $X$.
Based on the above construction, we have the following
facts.

\begin{fact}
\label{degree3rep}
Let $X$ be a weighted degree-$d$ phylogeny.  Let $T$ and $T'$ be two
degree-3 representaions of $X$.  Then,
\begin{itemize}
\item 
$\rsttdist (X,T) = 0$ and $\sttdist (T,X) = 0$, and
\item 
$\rsttdist(T,T')=0$.
\end{itemize}
\end{fact}

Also, STT operations on $X$ can
be ``simulated'' by a sequence of restricted-STT operations
on its degree-3 representation $T$
with the same cost. More precisely, we have the following lemma.

\begin{lemma}
\label{degree3sim}
Let $X$ be a weighted degree-$d$ phylogeny and
$X'$ be the phylogeny constructed from $X$ by one STT operation with
cost $c$. 
Let $T$ and $T'$ be the degree-$3$ representation of $X$ and $X'$, 
respectively.
Then, $T$ can be transformed to $T'$ using
a number of restricted-STT operations with the total cost $c$.
\end{lemma}
\begin{proof}
For each edge $e$ in $X$, there is a corresponding
edge $e'$ in $T$ such that they induce the same bipartition
of leaf labels. And if the nodes of $X$ are given
unique labels, then, for each node $u$ in $X$, there is a
corresponding node $u'$ in $T$ with the same label as $u$.
Now, we simulate an STT operation on $X$ in $T$ as follows.

If an STT operation moves a subtree $S$ (in $X$)
which is attached
to $u$ by an edge $e_1$ from $u$ to an edge $e_2$,
we simulate the operation in $T$ by moving the edge
$e_1'$ and its attached subtree to the edge $e_2'$.
On the other hand, if an STT operation moves the subtree $S$ and
$e_1$ from $u$ to an internal node $v$, there are two
cases. If $v$ is of degree 3, then let $e$ be any edge
attached to $v'$ in $T$, move $e_1'$ and its subtree to
$e$, and forming a new node $x$ on $e$ with $w((x,v'))=0$. 
Otherwise, $v$ must be of degree $k>3$, and
there must be an edge $\overline{e}$ attached to $v'$ (in $T$) such
that $\overline{e}$ induces a bipartition of leaf labels that
cannot be induced by any edge attached to $v$.
Intuitively, this is the new edge added when we transform
$u$ from degree $d$ to degree $3$. 
Then, we move $e_1'$ and its subtree to $\overline{e}$. It can be shown 
that $T'$
is a degree-3 representation of $X'$ and $\rsttdist(T,T')=c$.
\end{proof}

Now, we show that $\sttdist(X_1, X_2) = \rsttdist(T_1, T_2)$.

\begin{lemma}
\label{deg-dtodeg-3}
Let $X_1$ and $X_2$ be two weighted degree-$d$ phylogenies, where $d \geq 3$.
Let $T_1$ and $T_2$ be degree-$3$ representations of $X_1$ and
$X_2$, respectively.
Then, $\sttdist(X_1,X_2) = \rsttdist(T_1,T_2)$.
\end{lemma}
\begin{proof}
To transform $X_1$ into $X_2$ using STT operations, we can
first transform $X_1$ to $T_1$, then transform $T_1$ to $T_2$ using
restricted-STT operations, and finally transform $T_2$ back
to $X_2$. In other words,
$\sttdist(X_1,X_2) \leq \rsttdist(X_1,T_1)+\rsttdist(T_1,T_2)
+\sttdist(T_2,X_2)$.
By Fact \ref{degree3rep}, $\rsttdist(X_1,T_1)=0$ and
$\sttdist(T_2,X_2)=0$.
So, we have $\sttdist(X_1,X_2) \leq \rsttdist(T_1,T_2)$.

On the other hand, by Lemma \ref{degree3sim}, to tranform
$T_1$ into $T_2$ using restricted-STT operations, we can simulate
the transformation from $X_1$ to $X_2$ on $T_1$ to obtain $T'$
where $T'$ is a degree-3 representation of $X_2$. Then, we
tranform $T'$ to $T_2$.
By Fact \ref{degree3rep}, $\rsttdist(T',T_2)=0$. So,
$\rsttdist(T_1,T_2) \leq \sttdist(X_1,X_2)$. The lemma follows.
\end{proof}



By Lemma \ref{deg-dtodeg-3},
we have the following theorem.

\begin{theorem}
The weighted STT-d problem is equivalent to the weighted rSTT-3 problem.
\end{theorem}
\begin{proof}
Consider two weighted degree-d trees $X_1$ and $X_2$.
Let $R_1$ and $R_2$ be the degree-3 representations of
$X_1$ and $X_2$, respectively.  By Lemma~\ref{deg-dtodeg-3},
$\sttdist(X_1, X_2) = \rsttdist(R_1, R_2)$.
Thus, the weighted STT-d problem can be reduced to the weighted rSTT-3 problem.

Given two weighted
degree-3 trees $T_1$ and $T_2$,
$T_1$ and $T_2$ are its own degree-3 representations, respectively.
By Lemma~\ref{deg-dtodeg-3}, 
$\rsttdist(T_1, T_2) = \sttdist(T_1, T_2)$.
Hence, the weighted rSTT-3 problem can be reduced to the weighted STT-d problem.
\end{proof}

\subsection{An approximation algorithm for unweighted STT distance}
\label{sec-approx-STT-unweightedcase}
This section gives an approximation algorithm for
computing the STT distance between two unweighted phylogenies (the unweighted STT problem).
The approximation factor is $2d -4$, which is independent
of the number of leaves, $n$.

Given two phylogenies, we first define what a
{\it non-leaf-label-shared} edge is,
and give a lower bound on the STT distance between the
phylogenies based on the number of non-leaf-label-shared edges
in the phylogenies.

Let $T$ and $T'$ be two phylogenies with the same set of
leaf labels. An edge $e$ in $T$ is said to be
{\em leaf-label-shared} (w.r.t. $T'$),
if for some edge $e'$ in $T'$, $e$ and $e'$
induce the same partition of leaf labels; otherwise, $e$ is said to be 
\emph{non-leaf-label-shared}. 

\begin{lemma} \label{lowerbound}
Let $T$ and $T'$ be two degree-$d$  phylogenies with the same
set of leaf labels.  Let $b$ and $b'$ denote the number of 
non-leaf-label-shared edges in $T$ and $T'$, respectively.
Then $\sttdist(T,T') \geq \max(b,b')$.
\end{lemma}
\begin{proof}
By viewing an edge as a partition of leaves, a sequence of
STT operations with cost $k$ can create at
most $k$ new edges and delete at most  $k$
edges.  To transform $T$ to $T'$, we must either delete or create at
least $max(b, b')$ edges, because any non-leaf-label-shared
edge of one tree is not contained in another.
Thus, the STT distance is at least $max(b, b')$.
\end{proof}

Note that
STT operations are reversible in the sense that
if a tree $T_1$ can be transformed to $T_2$ using
a sequence $\sigma$ of STT operations, we can easily transform
$T_2$ to $T_1$ by reversing the operations in $\sigma$
with the same cost.
Based on the following lemma and the reversibility of STT
operations,
we will derive a linear time
approximation algorithm for computing the STT distance
between two phylogenies.

\begin{lemma}
\label{isomorphic}
{\rm (Theorem 4, \cite{Robinson:1981:CPT})}
Let $T$ and $T'$ be two unweighted degree-$d$ phylogenies with the same 
set of leaf labels.  If neither of them contains non-leaf-label-shared 
edges, then $T$ and $T'$ are isomorphic.
\end{lemma}

Figure \ref{STTapproxalg} details the approximation algorithm.
The basic idea is to transform
each phylogeny to one without non-leaf-label-shared edges using
STT operations.

\begin{figure}[htbp]
\fbox{
\begin{minipage}{.95\textwidth}
\noindent
\begin{itemize}
   \item[1.] Identify non-leaf-label-shared edges in $T$ and $T'$.
   \item[2.] Transform $T$ to $T_s$ by ``contracting" all
             non-leaf-label-shared edges using a sequence $\sigma_1$ of
             STT operations as follows:
             \begin{itemize}
             \item[2.1.] Partition the set of non-leaf-label-shared edges
             into groups such that if two edges are connected in $T$,
             they are in the same group.\\
             \item[2.2.] For each group, pick an internal node $x$ which
             attaches to one of the edges. For each non-leaf-label-shared
             edge $(x,y)$, let $k$ be the degree of $y$. By STT operations,
             we detach $(k-2)$ subtrees from $y$ and re-attach them
             to $x$. Then, $y$ becomes degree-$2$ and disappears. Repeat until all
             non-leaf-label-shared edges in the group are removed.
             \end{itemize}
   \item[3.] Repeat Step 2 to transform $T'$ to $T'_s$ using a
             sequence $\sigma_2$ of STT operations.
   \item[4.] Output the total cost $c$ of all STT operations in
             $\sigma_1$ and $\sigma_2$.
\end{itemize}
\end{minipage}
} 
\caption{An approximation algorithm for computing STT distance.}
\label{STTapproxalg}
\end{figure}

The following lemma analyses the approximation factor and the time
complexity of the algorithm.

\begin{lemma}
\label{performance}
Let $T$ and $T'$ be two degree-$d$ phylogenies with the same set of leaf
labels. Then, $\sttdist(T,T')$ can be approximated
within a factor of $2d-4$ in $O(n)$ time. 
\end{lemma}
\begin{proof}
We apply the approximation algorithm presented in Figure
\ref{STTapproxalg} to $T$ and $T'$.  Since the leaf labels are distinct
in $T$ ($T'$),  Step 1 can be done in $O(n)$ time \cite{Day:1985:OAC}. 
All other steps can be completed in $O(n)$ time, so the overall time
complexity is $O(n)$.

By Lemma \ref{isomorphic}, $T_s$ and $T'_s$ are isomorphic.  Thus,
by using the sequence of STT operations in $\sigma_1$ and reversing the
operations in $\sigma_2$, we can transform $T$ to $T'$ with cost $c$.

To determine the approximation factor, it remains to
bound the value of $c$.  Note that all the STT operations are
performed in Step 2, where we remove (contract) each 
the non-leaf-label-shared-edges.  The cost of STT operations for each
removal is at most $d-2$.  Thus, $c \leq b(d-2)+b'(d-2)$ where $b$ and
$b'$ are  the number of non-leaf-label-shared-edges in $T$ and $T'$,
respectively.  By Lemma \ref{lowerbound}, $c \leq
(2d-4)\cdot\mbox{max}(b,b') \leq (2d-4)\sttdist(T,T')$, 
so the lemma follows.
\end{proof}

\section{Unweighted degree-$d$ phylogenies and NP-hardness}
\label{sec-STT-nphard}
This section studies the computational complexity for
computing the STT distance between two unweighted degree-$d$
phylogenies. We prove the NP-hardness of a slightly more
general problem. We prove that the problem
of computing the STT distance between two unweighted trees with
leaves labeled by possibly non-distinct labels
is NP-hard. Our result also implies
the NP-hardness of the weighted version of this problem,
which was first proven in \cite{DasGupta:1999:OLC}.

We consider the following decision problem.
Given two degree-$d$ unrooted trees $T$ and $T'$ and an integer $t$,
the problem is to determine whether $\sttdist(T, T') \leq t$.
We show that this decision problem is NP-hard by reducing
the Exact Cover by 3-Sets (X3C) problem \cite{Garey:1979:CIG} to it.
The X3C problem is defined as follows.
Given a set $S = \{s_1, s_2, \ldots, s_{3q} \}$ for some integer $q$
and a collection ${\cal C}$ = $C_1, C_2, \ldots, C_n$ of subsets of
$S$ where $C_i = \{s_{i_1}, s_{i_2}, s_{i_3}\}$
and $\cup_{i=1}^n C_i = S$, 
determine whether ${\cal C}$ contains an exact cover
for $S$, that is, a sub-collection ${\cal D} = D_1, D_2, \ldots, D_q$
of ${\cal C}$ such that $\cup_{j=1}^{q}{D_j} = S$.
Given an instance of X3C problem, we construct two degree-$d$
trees $T$ and $T'$ where $d=4n-q$ as follows.

\begin{figure}[htbp]
   \begin{center}   
   \psfrag{(a)}{{\Large (a)}} \psfrag{Ci}{{\large $C_i$}} 
   \psfrag{si1}{{\Large $s_{i_1}$}} 
   \psfrag{si2}{{\Large $s_{i_2}$}}
   \psfrag{si3}{{\Large $s_{i_3}$}}
   \psfrag{(b)}{{\Large (b)}}
   \psfrag{short_arm}{{\large short arm}}
   \psfrag{long_arm}{{\large long arm}} 
   \psfrag{3_short_arms}{{\large $3$ short arms}} 
   \psfrag{x0}{{\Large $x_0$}} \psfrag{x1}{{\Large $x_1$}} 
   \psfrag{xn2-2}{{\Large $x_{n^2-2}$}}
   \psfrag{xn2-1}{{\Large $x_{n^2-1}$}} 
   \psfrag{(c)_The_tree_T}{{\Large (c) The tree $T$}}
   \psfrag{n_long_arms}{{\large $n$ long arms}}
   \psfrag{M}{{\large $M$}}
   \psfrag{C1}{{\large $C_1$}} \psfrag{C2}{{\large $C_2$}}
   \psfrag{Cn-1}{{\large $C_{n-1}$}}
   \psfrag{Cn}{{\large $C_n$}}
   \psfrag{(d)_The_tree_T'}{{\Large (d) The tree $T'$}}
   \psfrag{N}{{\large $N$}}
   \psfrag{n-q_long_arms}{{\large $n-q$ long arms}}
   \psfrag{3q_short_arms}{{\large $3q$ short arms}}
   \psfrag{3n-3q}{{\large $3n-3q$}}
   \psfrag{remaining_leaves}{{\large remaining leaves}}
   \psfrag{s1}{{\Large $s_1$}}  \psfrag{s2}{{\Large $s_2$}}
   \psfrag{s3q-1}{{\Large $s_{3q-1}$}} \psfrag{s3q}{{\Large $s_{3q}$}}
   \psfrag{=}{{\Large $=$}}
   \resizebox{6.2in}{!}{\includegraphics{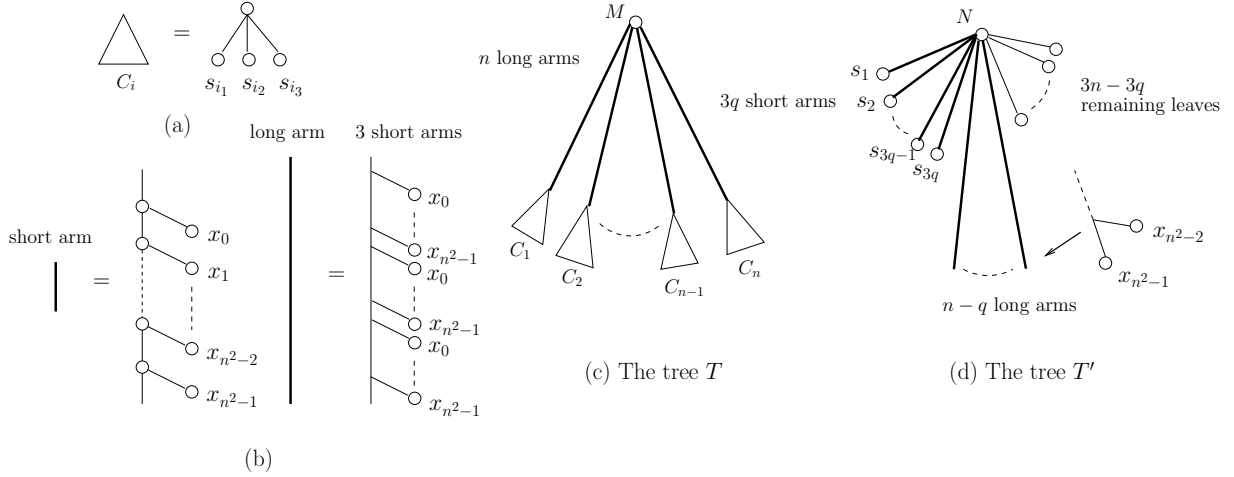}}
   \end{center}
\caption{Trees $T$ and $T'$.} 
\label{NP-Complete_fig}
\end{figure}

\vspace{5pt}
\noindent
{\it Construction of $T$}: For each $C_i = \{s_{i_1}, s_{i_2}, s_{i_3}\}$,
we construct a subtree with three leaves labeled as $s_{i_1}, s_{i_2},
s_{i_3}$, respectively (see Figure \ref{NP-Complete_fig}(a)).
In $T$, each of these subtrees is attached to a long arm where a long arm
is made up of three short arms.
Each short arm is a path with $n^2$ leaves labeled as
$x_0, x_1, \ldots, x_{n^2-1}$ hanging on the path
(see Figure \ref{NP-Complete_fig}(b)). In other words, besides $3q$ $s_i$
labels, we create $n^2$ unique $x_i$ labels. All long arms meet
at a single node $M$ (see Figure \ref{NP-Complete_fig}(c)).

\noindent
{\it Construction of $T'$}:
We construct $3q$ leaves with labels $s_1, \ldots, s_{3q}$, respectively.
Each of these leaves is attached to a short arm
as shown in Figure \ref{NP-Complete_fig}(d).
The short arms meet at a node $N$. Besides the short arms, there
are $n-q$ long arms attached to $N$. To make $T'$ have the same
multi-set of leaf labels as $T$, we create additional leaves with
appropriate labels and attach these leaves to $N$ directly.

Let $t = 3n^3 + 6n$.
We show that $\sttdist(T, T') \leq t$ if and only if
$S$ has an exact cover. If $S$ has an exact cover,
there exists $q$ long arms in $T$ such that the set of
all $s_i$ leaves at the end of these long arms is equal
to $S$. To transform $T$ to $T'$, 
basically, each of these long arms is transformed
into 3 short arms with one $s_i$ leaf attached
at the end. For the other long arms of $T$, we move up
all $s_i$ leaves to $M$. The whole proceduce can be
done using STT operations of cost at most $t$.
The detail steps are given below (c.f. \cite{DasGupta:1999:OLC}).
Although $T$ is an unrooted tree, for ease of description,
if a STT operation moves a subtree towards $M$, we say that
it {\it moves the subtree up} the tree, otherwise, we say
that it {\it moves the subtree down} the tree.

\begin{figure}[htb]
   \begin{center}   
   \psfrag{(a)}{{\Large (a)}} \psfrag{(b)}{{\Large (b)}} 
   \psfrag{x0}{{\Large $x_0$}} \psfrag{xn2-1}{{\Large $x_{n^2-1}$}} 
   \psfrag{sp}{{\Large $s_p$}} \psfrag{sr}{{\Large $s_r$}} 
   \psfrag{sq}{{\Large $s_q$}} 
   \psfrag{M}{{\Large $M$}} \psfrag{P}{{\Large $P$}} 
   \psfrag{R}{{\Large $R$}} 
   \psfrag{=>}{$\Rightarrow$}      
   \resizebox{6.2in}{!}{\includegraphics{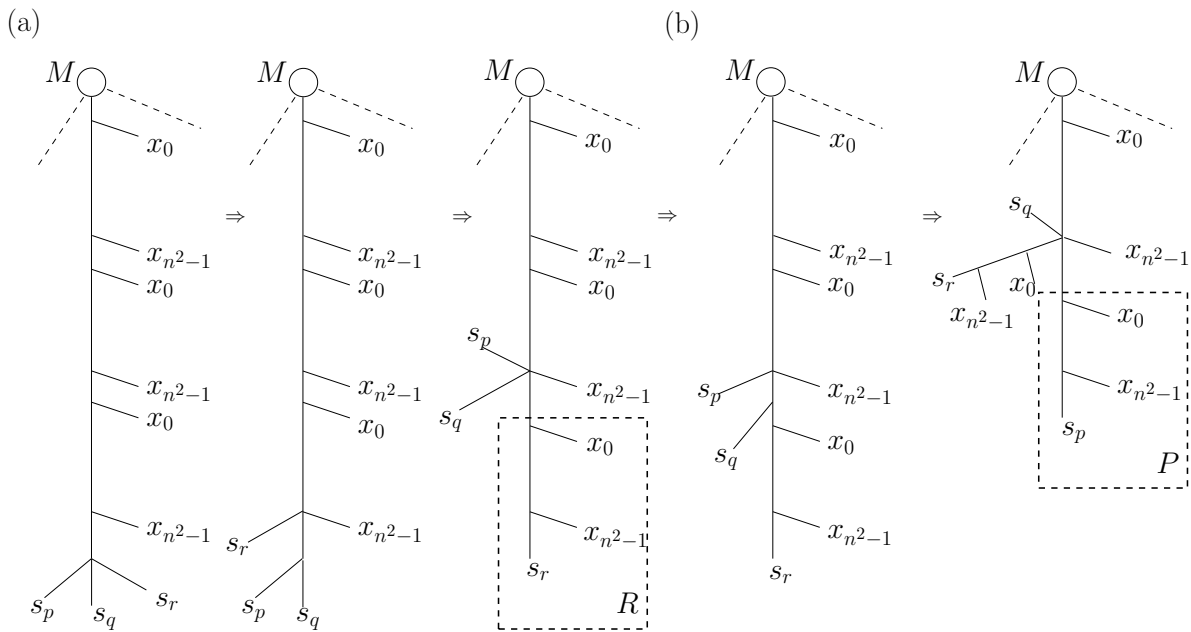}}
   \end{center}
\caption{Transforming a long arm in $T$ to three short arms in $T'$.} 
\label{if-part}
\end{figure}

Without loss of generality, let $\{C_1, C_2, \ldots, C_q\}$ be
an exact cover of $S$. There are two cases.
\begin{description}
\item[Case 1.] For the long arm corresponding to $C_i$ where $1 \leq i \leq q$,
      let $C_i = \{s_p, s_q, s_r\}$. We first move the two leaves
      with labels $s_p$ and $s_q$ up $n^2+1$ edges (see Figure
      \ref{if-part}(a)) using
      STT operations of cost $n^2+2$. Note that the subtree $R$ (see
      Figure~\ref{if-part}(a)) will be a short arm in $T'$. The next step
      is to move the leave with label $s_q$ and $R$
      up $n^2+1$ edges (see Figure \ref{if-part}(b))
      using STT operations of cost $n^2+2$. Note that the subtree
      $P$ will be another short arm in $T'$. The last step is to
      move both $P$ and $R$ to $M$ and make them attach to $M$ directly.
      This requires STT operations of cost $n^2+2$.
      The total cost for each long arm is $3n^2+6$. So, the
      total cost of to transform these long arms into short
      arms is $3n^2q + 6q$.

\item[Case 2.] For each of the remaining long arms, move the subtree
      containing the three $s_i$ leaves up to $M$ using
      STT operations of cost $3n^2$, then using two
      more STT operations to make each $s_i$ leaf attached
      to $M$ directly. The total cost of STT operations
      for these long arms is $(3n^2 + 2)(n-q)$.
\end{description}

Thus, we can transform $T$ to $T'$
using STT operations with total cost at most
$3n^3+6n$. The correctness of the if-part is
established.
In the following, we focus on showing the correctness of the
only-if part. That is, we show that 
if $\sttdist(T, T') \leq 3n^3+6n$, then there
is an exact cover of $S$.

\begin{observation} \label{obs-unit-cost}
For any STT operation of cost $k$, we can decompose it into 
$k$ STT operations of which each has cost one.
\end{observation}

From this point onwards, we regard each STT operation as of unit cost unless
otherwise stated. In other words, each STT operation will move a
subtree together with the edge attached to it
from one end of an edge to the middle or the other end of
the same edge.
To prove the only-if part, for a given sequence of STT operations
that transform $T$ to $T'$, we identify a set of {\it effective}
STT operations. We show that each of these operations can
be characterized by a unique edge (called an {\it upward} edge)
in $T$. If the total number of STT operations
is small (i.e., $\leq 3n^3+6n$), then the number of effective
operations must be large, that is, there must be a lot of upward
edges and this implies the existence of
an exact cover for $S$.
We first give some preliminary definitions and
concepts in Section \ref{only-if-def-sec}.
We then give a lower bound on the number of upward edges
in Section \ref{uparrow-lb-sec}. The proof of the only-if
part is given in Section \ref{only-if-proof-sec}.

\subsection{Definitions and concepts} \label{only-if-def-sec}

Let $F$ be a given sequence of STT operations of total
cost at most $3n^3+6n$ that
transforms $T$ to $T'$. By tracing the operations, there
is a one-to-one correspondence of the leaves in $T$ to those
in $T'$. We can relabel the leaves of $T$ by giving an
extra index to the leaves with same labels. For example,
we can use $x_{0,1}, x_{0,2}, \ldots$ to distinguish leaves with label
$x_0$. The leaves in $T'$ can be relabeled according to the new
labels of the corresponding leaves in $T$.
After the relabeling, we can regard leaves in $T$ (or $T'$)
as having distinct labels. Since the index $j$ in $x_{i,j}$ is not
important, so we will refer to any particular $x_{i,j}$ simply
by $x_i$ in the following. Similarly for $s_i$ leaves, we relabel them
and refer to them using the same approach. In other words,
we can now assume that all leaf labels are distinct with respect
to $F$.

Each edge $e$ in $T$ induces a bipartition of leaves, denoted as $b_e$.
After the relabeling, these bipartitions are unique.
Let ${\cal B}=\{b_1,b_2,\ldots,b_{3n^3+n}\}$ and
${\cal B}'=\{b'_1, b'_2, \ldots, b'_{3n^3-n+q}\}$ be the
the set of bipartitions
induced by the internal edges in $T$ and $T'$, respectively.
Note that ${\cal B} \cap {\cal B}' = \emptyset$.

\begin{figure}[htb]
\centerline{\psfig{figure=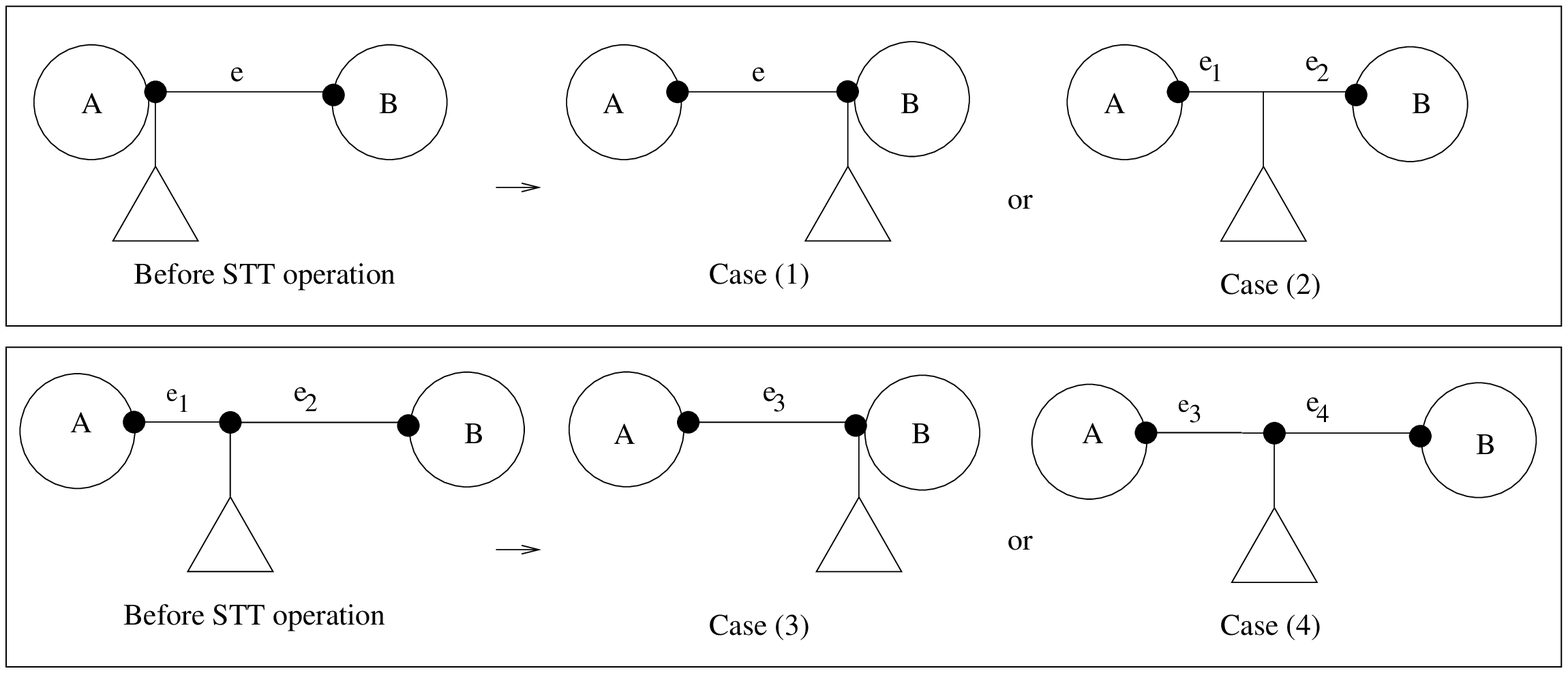,height=2.5in,width=5in}}
\caption{Possible STT operations.} 
\label{stt-operation}
\end{figure}

Let $T_i$ be the resulting tree after performing
the $i$th STT operation in $F$. Let ${\cal B}^i$ be the set of
bipartitions induced by the internal edges in $T_i$.
For any $i$, since the leaves are uniquely relabeled,
the bipartitions in ${\cal B}^i$ are unique.
Let $op$ be an STT operation in $F$ that transforms $T_i$ to $T_{i+1}$.
After the operation, we either
(1) delete a bipartition $b \in {\cal B}^i$ and create a new bipartition
$b' \in {\cal B}^{i+1}$;
or (2) create a new bipartition without deleting any bipartiton
in ${\cal B}^i$; or
(3) delete a bipartition in ${\cal B}^i$; or
(4) do not change the set of bipartitions.
See Figure \ref{stt-operation} for
all these STT operations. 

An STT operation $op$ is called an {\it effective
operation} if $op$ deletes a bipartition $b \in {\cal B}^i$ and creates
a new bipartition $b' \in {\cal B}^{i+1}$ (Case 1)
such that
$b$ in ${\cal B}$, $b'$ in ${\cal B}'$ and no subsequent
STT operations in $F$ delete $b'$.
The unique edge $e$ in $T$ which induces $b$ is called an {\it effective
edge} and we say that $e$ maps $\tilde{e}$ if $\tilde{e}$
induces $b'$ in $T'$.  Finally, an STT operation which is not 
effective is called an {\it ineffective operation}.

Note that each effective operation $op$ can be characterized by
the pair of edges $(e, \tilde{e})$.
The following lemma
gives the bounds for the numbers of effective and ineffective operations.

\begin{lemma} \label{NP-hard:counting}
If $~\sttdist(T,T') \leq 3n^3+6n$, then
{\tt (i)} the number of effective operations is at least $3n^3 - 6n$ and  
{\tt (ii)} the number of ineffective operations is at most $12n$.
\end{lemma}
\begin{proof}
Since ${\cal B} \cap {\cal B}' = \emptyset$, there must be STT operations that
delete all bipartitions in ${\cal B}$ and create the bipartitions in
${\cal B}'$. $|{\cal B}| = 3n^3+n$ and $|{\cal B}'| = 3n^3-n+q \geq 3n^3-n$.
Since the total number of
STT operations is at most $3n^3+6n$, 
the number of effective operations, each of which deletes a bipartition in
${\cal B}$ and creates a bipartition in ${\cal B}'$, must be at least
$(3n^3+n)+(3n^3-n) - (3n^3+6n) =
3n^3-6n$.  The number of ineffective operations must be at most
$(3n^3+6n) - (3n^3-6n) = 12n$.
\end{proof}

To help the analysis, we further classify effective edges in $T$.
Let $e$ be an effective edge. Let $op$ be the
effective operation that deletes the bipartition induced by $e$.
If $op$ moves a subtree up the tree, $e$ is called an {\it upward
effective edge}. Otherwise, it is called a {\it downward effective
edge}.

The number of upward effective edges is critical to the proof
of the only-if part. In the next section, we will show
that the number of upward effective edges
will be large if the total number of STT operations is
at most $3n^3+6n$.

\subsection{Lower bound on the number of upward effective edges}
\label{uparrow-lb-sec}
Consider some large enough $n$.
This section shows that
if {\tt STTdist}$(T,T') \leq 3n^3 + 6n$, then
every long arm in $T$ contains more than $2n^2$ upward effective
edges.

We first show that the number of downward effective edges is at most
$89n$. We classify the downward effective edges into the following
three groups. For each downward effective edge, there is a corresponding
STT operation that moves a subtree $S$ and its attached edge $e$,
called the {\it carrying edge}, from one end of some edge to the
other end. The subtree $S$ is said to be carried by the edge $e$.
Let $b_e$ be the bipartition induced by the carrying edge $e$.
We count the number of downward effective edges in each of the
following groups:
(a) $e$ is an external edge or $b_e \in {\cal B}'$,
(b) $b_e \in {\cal B}$, and
(c) $e$ is not an external edge and $b_e \not \in {\cal B} \cup {\cal B}'$.

\vspace{3pt}
\noindent
Case (a). In this case, the subtree carried by
$e$ must have exactly $0, 1, 3n-1$ or $3n$ leaves with $s_i$ labels
while the biparition induced by the
downward effective edge has 3 leaves with $s_i$ labels on one
side and $3n-3$ leaves with $s_i$ labels
on the other side. By checking all cases, it is not
possible to produce a biparition in ${\cal B}'$ using the
corresponding carrying edge.
The number of downward effective edges in this group is 0.

\vspace{3pt}
\noindent
Case (b). In this case, since $b_e$ is in ${\cal B}$,
let $e_i$ be the edge in $T$ that induces the same $b_e$.
Let the corresponding downward effective edge be $e_j$
and let $e_j$ maps $\tilde{e_j}$.
By checking all possible cases, we know that both $b_{e_i}$
and $b_{e_j}$ must contain the same set of 3 leaves with
$s_i$ labels in one side of the bipartition.
From the structure of $T$, we know that $e_i$ and $e_j$
must be in the same long arm $A$. Since $e_j$ is
a downward effective edge, $e_i$ is closer to $M$ (the
node where other long arms meet).

Consider $A$ and denote its edges by
$e_0, e_1, \ldots, e_{3n^2-1}$ ordered from $M$.
The set of leaves $L$ between $e_i$ and $e_j$ in $T$ contains
leaves of labels
$x_{i \mod n^2}$, $x_{(i+1) \mod n^2}$, $\ldots$,
$x_{(j-1)\mod n^2}$.
It can be easily verified that after the STT operation,
one side of $b_{\tilde{e_j}}$ will contain leaves of
these $x_i$ labels only. And this partition has the
following property. If it has $m$ leaves of
label $x_p$, it 
must contain at least $m$ leaves of label
$x_{p+1}$, $\ldots$, $x_{n^2-1}$. Otherwise, it cannot
induce a bipartition in $T'$. Therefore, we
deduce that if $j \neq n^2, 2n^2$, then
$i = j \mod n^2$.

When $i = j \mod n^2$, we know that $L$ contains same number of
leaves with label $x_0, x_1, \ldots, x_{n^2-1}$. 
In $T'$, the number of bipartitions in ${\cal B}'$ which have
a side with equal number of leaves with labels $x_0, \ldots, x_{n^2-1}$
is at most $3n$. Since each effective STT operation will be
corresponding to a unique edge in $T'$. So, there are at most
$3n$ downward effective edges when $j \neq n^2, 2n^2$.
And if we include the case when $j = n^2, 2n^2$, then 
the total number of downward effective edges in this group is at most
$5n$.

\vspace{3pt}
\noindent
Case (c). In this case, the bipartition induced by
the carrying edge $e$ is not in ${\cal B}$ or ${\cal B}'$.
First, we have the following observation.

\begin{observation} \label{obs-2}
Let $e$ and $e'$ be downward effective edges such that
their corresponding carrying edges induce the same
bipartition of leaf labels.
Then $e$ and $e'$ are in the same long arm in $T$.
\end{observation}

The next lemma shows
that there are at most 7 downward
effective edges
in the same long arm in $T$ such that the corresponding
carrying edges induce the same bipartition of leaf labels.

\begin{lemma} \label{lem-3.8}
For each carrying edge $e$, there are at most 7 downward effective edges
induced by $e$. 
\end{lemma}
\begin{proof}
Fixed a long arm in $T$ and denote its edges by
$e_0, e_1, \ldots, e_{3n^2}$ from $M$.
Let $e_{i_1}$, $e_{i_2}$, $\ldots$,
$e_{i_m}$ ($i_1 < i_2 < \dots < i_m$) be $m$
downward effective edges induced by $e$.

For $k = 1, 2, \ldots, m-1$,
let $L_k$ be the set of leaves betweeen $e_{i_k}$ and $e_{i_{k+1}}$.
Thus $L_k = \{ x_{i_k \mod n^2}, \ldots, x_{(i_{k+1}-1) \mod n^2} \}$.
We claim that $L_k \cup L_{k+1}$ contains at least one $x_{n^2-1}$ leaf 
for all $k=1, \ldots, m-2$.  The claim can be proved as follows.

By contrary, assume $L_k \cup L_{k+1}$ does not contains $x_{n^2-1}$.
Then, the sets $L_k$ and $L_{k+1}$ should
be $\{ x_a, \ldots, x_{b-1} \}$ and $\{ x_b, \ldots, x_c \}$
respectively where $a = i_k \mod n^2$, $b = i_{k+1} \mod n^2$,
$c = (i_{k+2}-1) \mod n^2$, and $0 \leq a \leq b \leq c < n^2$.

Let $e_{i_k}$, $e_{i_{k+1}}$, $e_{i_{k+2}}$ map to $\tilde{e}_{i_k}$,
$\tilde{e}_{i_{k+1}}$, and $\tilde{e}_{i_{k+2}}$, respectively.
Let $R$ be the partition of $b_{e_{i_k}}$ which contains more than 3 
$s_i$ leaves. Let $R'$ be the set of leaves in $R$ excluding the leaves in the 
subtree carried by $e$. 
Note that $R'$ should have less than or equal to 1 $s_i$ leave.
Otherwise, both partitions of $b_{\tilde{e}_{i_k}}$ in $T'$ contains
more than 1 $s_i$ leaf, which is impossible.
Thus, $R'$ contains all leaves below $\tilde{e}_{i_k}$
of some arm $A'$ in $T'$.

Using the same argument for $\tilde{e}_{i_{k+1}}$ and
$\tilde{e}_{i_{k+2}}$, we can show that
$R' \cup L_k$ and $R' \cup L_k \cup L_{k+1}$ contains all leaves
below $\tilde{e}_{i_{k+1}}$ and $\tilde{e}_{i_{k+2}}$
respectively of $A'$ in $T'$.
Recall that $x_a \in L_k$ and $x_c \in L_{k+1}$.
By construction of $T'$, $c < a$.
We arrive at contradiction and the claim follows.

%
%
%
%

Since there are only 3 $x_{n^2-1}$ leaves in each arm of $T$, the above
claim implies that there are at most 7 downward effective edges induced by 
$e$.
\end{proof}
By Observation 2 and Lemma~\ref{lem-3.8}, each carrying edge corresponds to at 
most 7 downward effective edges in $T$. 
Each such carrying edge $e$ requires a distinct ineffective STT operation 
to delete the corresponding bipartition
$b_e \not \in {\cal B} \cup {\cal B}'$, therefore, the number of
downward effective edges in this group is at most $7*12n=84n$, since
by Lemma \ref{NP-hard:counting},
there are at most $12n$ ineffective STT operations.

Summing up the number of possible downward effective edges, we have
the following lemma.

\begin{lemma} \label{lem-downarrow}
The number of downward effective edges is at most $89n$.
\end{lemma}
\begin{proof}
Recall the classification of downward effective edges in this subsection.
The number of such edges in cases {\rm (a)}, {\rm (b)} and {\rm (c)} are
at most $0$, $5n$ and $84n$, respectively.  Thus, the total number is at
most $89n$. 
\end{proof}

Combining lemmas \ref{NP-hard:counting} and \ref{lem-downarrow},
we have the following.

\begin{lemma} \label{lem-part1}
If $~\sttdist(T, T') \leq 3n^3+6n$, then
$T$ has at least $3n^3 - 95n$ upward effective edges.
\end{lemma}
\begin{proof}
The number of upward effective edges
\[
\begin{array}{ll}
= & \mbox{number of effective operations} - \mbox{number of downward effective edges} \\
\geq & \mbox{number of effective operations}-89n \mbox{ (by Lemma~\ref{lem-downarrow})} \\ 
\geq & 3n^3 - 6n - 89n \\
= & 3n^3 - 95n\mbox{.}
\end{array}
\]
Thus, Lemma~\ref{lem-part1} follows.
\end{proof}

Based on Lemma \ref{lem-part1},
we have the following corollary.

\begin{corollary} \label{cor-uparrow}
For large enough $n$ ($n \geq 96$),
if $~\sttdist(T, T') \leq 3n^3+6n$, then
every long arm in $T$ contains more than $2n^2$ upward effective edges.
\end{corollary}
\begin{proof}
We prove the corollary by contradiction.
Suppose one of the long arms contains at most $2n^2$ upward effective edges,
then the number of upward effective edges in $T$ at most
$(3n^2+1)(n-1)+ 2n^2 = 3n^3-n^2+n-1$.  
However, since $\sttdist(T, T') \leq 3n^3+6n$, by 
Lemma~\ref{lem-part1}, $T$ should has at least $3n^3-95n$ upward effective
edges.  We arrive at a contradiction and the corollary follows.
\end{proof}

\subsection{Proof of the only-if part}
\label{only-if-proof-sec}

Now, we prove the only-if part.

\begin{lemma} \label{lem-leaf}
If an upward effective edge $e$ in a long arm $A$ in $T$
maps to an edge $\tilde{e}$ in an arm $A'$ in $T'$,
then the leaves below $\tilde{e}$ in $A'$ are originally
the leaves below $e$ in $A$.
\end{lemma}
\begin{proof}
Consider a particular leaf $\ell$ which is below $\tilde{e}$ in $A'$.
To prove by contradiction, suppose $\ell$ is originally not a leave below $e$ in $A$,
i.e., $\ell$ originally appears on the bipartition of $e$ 
with more than $3n-3$ $s_i$ leaves.
Since $e$ is an upward effective edge, this means that
after $\tilde{e}$ is created,
$\ell$ will remain on the bipartition of
$\tilde{e}$ with more than $3n-3$ $s_i$ leaves, which is
a contradiction.
\end{proof}

\begin{lemma} \label{lem-uparrow-distinct}
Upward effective edges from distinct long arms in $T$ cannot map edges
in the same arm in $T'$.
\end{lemma}
\begin{proof}
Let $e_1$ and $e_2$ be upward effective edges from distinct long arms
$A_1$ and $A_2$ in $T$.  
Suppose on contrary that the edges $\tilde{e_1}$ and $\tilde{e_2}$,
which are mapped from $e_1$ and $e_2$,
are in the same arm, say $A'$, in $T'$.
Consider the unique leaf $x_{n^2-1}$ at the bottom of $A'$ in $T'$.
By Lemma~\ref{lem-leaf}, this unique leaf
should be below $e_1$ in $A_1$ in $T$.
Similarly, we can show that the unique leaf should be below $e_2$ in $A_2$ in $T$.
The uniqueness of the leaf implies that $e_1$ and 
$e_2$ are in the same arm.  Thus, contradiction occurs and the lemma 
follows.
\end{proof}

Based on Corollary \ref{cor-uparrow} and 
Lemmas \ref{lem-leaf} and \ref{lem-uparrow-distinct}, the short arms in $T'$ can
be divided into groups of 3. Each group corresponds to one long
arm in $T$ where the leaves of the same group are exactly those
leaves of the corresponding long arm. Thus,
we have the following theorem.

\begin{theorem}
If $\sttdist(T, T') \leq 3n^3+6n$, then $S$ has an exact
cover.
\end{theorem}
\begin{proof}
Suppose $\sttdist(T, T') \leq 3n^3+6n$.
By Corollary \ref{cor-uparrow},
every long arm in $T$ contains more than $2n^2$ upward effective edges.
By Lemma~\ref{lem-uparrow-distinct}, there exist at most $n-q$ long
arms in $T$ whose upward effective edges can create edges in the $n-q$
long arms in $T'$.

Let $R$ be the set of remaining (at least $q$) long arms in $T$.
Since each long arm in $R$ contains more than $2n^2$ upward effective
edges
and each short arm in $T'$ contains $n^2$ edges,
the upward effective edges on each long arm in $R$
create edges in at least $3$ short arms in $T'$.
As there are $3q$ short arms in $T'$, 
we conclude that $R$ contains exactly $q$ long arms
and each short arm in $T'$ has at least $1$ edge
created from some upward effective edge in long arm in $R$.
By Lemma~\ref{lem-leaf}, every $s_i$ at the bottom of the short arms of $T'$
should appear in the long arms in $R$.
It implies that the set of 
leaves with label $s_i$ at the bottom of the $q$ long arms must be equal 
to those at the bottom of the $3q$ short arms.  In other words, there is an 
exact cover of $S$.  This concludes the correctness of the only-if part.
\end{proof}

\section{Conclusions}
In this paper, we have explored different metrics for comparing phylogenies.
We have devised an $O(n\log n)$-time algorithm for computing the non-shared
edges, which in turn reduces the time complexity of existing approximation
algorithms for NNI distance from $O(n^2)$ to $O(n\log n)$.   On the other
hand, we have extended the study of STT distance to general degree-$d$ trees.
For weighted case, we have shown that the STT distance of two degree-$d$ trees
is equivalent to the restricted STT distance of two degree-3 trees.
For unweighted case, we have given an algorithm that approximates STT distance
within a factor of $2d-4$.  Also, we have shown that computing STT distance
between two non-distinctly labeled trees is NP-hard.

For future work, we would like to know whether there exists a linear-time
algorithm for computing the non-shared edges, and whether computing STT
distance between two distinctly labeled trees is NP-hard.

\bibliographystyle{plain}

\end{document}